\documentclass[aps,prb,twocolumn,showpacs,floatfix,superscriptaddress,amssymb,amsmath]{revtex4}
\usepackage{graphicx}
\usepackage{dcolumn}
\usepackage{bm}

\newcommand{\be}{\begin{equation}}
\newcommand{\ee}{\end{equation}}
\newcommand{\bea}{\begin{eqnarray}}
\newcommand{\eea}{\end{eqnarray}}

\newcommand{\s}{\sigma}

\newcommand{\ri}{\mbox{i}}

\newcommand{\ba}{\begin{array}}
\newcommand{\ea}{\end{array}}
\def\nn{\nonumber\\}

\def\up{\uparrow}
\def\down{\downarrow}

\begin{document}

\title{Landau level mixing by full spin-orbit interactions}

\author{M. Zarea and S. E. Ulloa}
\affiliation{Department of Physics and Astronomy, and Nanoscale
and Quantum Phenomena Institute, \\Ohio University, Athens, Ohio
45701-2979}

\date{\today}

\begin{abstract}
We study a two-dimensional electron gas in a perpendicular
magnetic field in the presence of both Rashba and Dresselhaus
spin-orbit interactions. 
Using a Bogoliubov transformation we are able to write
 an approximate formula for the Landau levels, thanks to
the simpler form of the resulting Hamiltonian.
The exact numerical
calculation of the energy levels, is also made simpler by our
formulation. 
The approximate formula and the exact numerical results
 show excellent agreement for
typical semiconductors, especially at high magnetic fields. 
We also show how effective Zeeman coupling is modified by spin-orbit
interactions.
\end{abstract}

 \keywords{}

\maketitle

\section{Introduction}

The manipulation of the spin of  charge carriers in
semiconductors, spintronics, has attracted  increasing interest in
recent years. In the paradigmatic Datta-Das spin
transistor,\cite{Datta} the spin of the electron passing through
the device is controlled by the Rashba spin-orbit (SO)
interaction,\cite{Rashba} which in turn can be varied by the
application of gate voltages.  The Rashba interaction stems from
the structural inversion asymmetry (SIA) introduced by a
heterojunction or by surface or external fields. In semiconductors
with narrower  energy gap (InGaAs, AlGaAs), this effect is
expected to be stronger. It has been shown experimentally that the
Rashba spin-orbit interaction can be modified up to $50\%$ by
external gate voltages.\cite{Miller,Nitta}

In addition to the Rashba coupling there is also a
material-intrinsic Dresselhaus spin-orbit interaction. This
originates from the bulk inversion asymmetry (BIA) of the crystal,
and can be relatively large in semiconductors like
InSb/InAlSb.\cite{Dress}

Both Rashba and Dresselhaus interactions contribute to the
spin-dependent splitting of the band structure of the host
material, leading to  dramatic spin-dependent phenomena for
electrons or holes in semiconductors: Effects on the charge and
magnetic transport, spin relaxation, and spin-Hall conductivity
have been recently studied.\cite{Lang,Heida,Averkiev,
Kainz,Sinova,Syn} A drift-diffusive  transistor, in contrast to the
ballistic Datta-Das device, has been proposed to be more robust by
Schliemann {\em et al}.\cite{Loss} for the case when the two SO
interactions have the same strength. The experimental observation
of the spin-galvanic effect and weak localization effects have
increased the interest in understanding the interplay between
different SO terms.\cite{Pikus} Photocurrent measurements have
been used to obtain the ratio of Rashba and Dresselhaus
coefficients.\cite{Ganichev} Beautiful optical measurements of SO
effects have also been performed recently in strained
semiconductors.\cite{Kato}

It is well known that the determination of the eigenvalues 
and eigenstates of the 
system is crucial for the calculation of a number of important 
physical properties of the system. In the case of a two-dimensional
electron gas (2DEG)   in
a  perpendicular magnetic field in the presence of both SO
interactions, few solvable cases have been analyzed in the literature
(we devote the next section to review them). However, a comprehensive
description of the more general case, in which both Rashba and Dresselhaus
interactions are present with arbitrary strength, would be interesting.

In this work, we address the problem of SO coupling effects on the
Landau level structure of a 2DEG in
a strong perpendicular field. We find an excellent approximate
expression for the spinor Landau levels in the most general case
of different Rashba and Dresselhaus interactions. We compare our
approach to numerical results, made easier and recursively exact
by our formulation of the problem. We further study the effective
Zeeman g-factor of the system, and spin-orbit coupling is found to
enhance or suppress the Zeeman splitting, depending crucially on
material parameters and gate voltages. This behavior may be found
useful in the characterization of spin-filter\cite{Hauson} and
spin-polarized currents in two-dimensional systems. \cite{Potok}

In the next section we review the exactly solvable cases of SO
coupling in a field, where either Rashba or Dresselhaus coupling
is present. The third section is dedicated to the general case in
which both terms are present. Using a Bogoliubov transformation we
transform the  Hamiltonian of the 2DEG in a perpendicular magnetic
field, and in the presence  of both Rashba and Dresselhaus terms,
to the 2DEG with only an effective Rashba interacting term with
modified strength. The specific form of the interacting term
allows for the derivation of the numerical exact level structure
of the model. Our approximate result is obtained and shown to be
increasingly accurate for higher magnetic fields or weaker SO
interaction. The last section contains typical results for
different materials and discussion.

\section{Solvable cases}

The Hamiltonian of  2D electrons with effective mass $m$ and
Zeeman coupling $g$ in a perpendicular magnetic field $B\hat z$ is
\be H_0=\frac{P_x^2}{2m}+\frac{P_y^2}{2m}-\frac{g\mu_BB}{2}\s^z,
\ee where $\mu_B$ is the Bohr magneton, $\s^z$ is the Pauli
matrix, and $ P=\vec{p}+\frac{e}{c}\vec{A}$ is the kinetic
momentum. It is well known that this Hamiltonian can be written as
a simple harmonic oscillator describing the Landau levels:
 \be
H_0=\hbar\omega_c(a^{\dag}a+\frac{1}{2})+\hbar\omega_c\xi\s^z,
 \ee
in which $a=(P_y+\ri P_x)/\sqrt{2m\hbar\omega_c}$,
$\omega_c=eB/mc$ and $\xi={-g\mu_Bmc/2\hbar e}$.

In this system, both Rashba and the linear Dresselhaus
interactions have a simple form given by
 \bea
H_R=\frac{\alpha_0}{\hbar}(P_y\s^x-P_x\s^y) &=&
\hbar\omega_c\alpha\left(\ba{cc} 0 &  a\\ a^{\dag} &0 \ea
\right)\nn H_D=\frac{\beta_0}{\hbar}(P_x\s^x-P_y\s^y) &=&
\hbar\omega_c\beta\left(\ba{cc} 0 & \ri a^{\dag}\\-\ri a &0 \ea
\right).\label{SO-Ha}
 \eea
Here we have introduced the dimensionless parameters
$\alpha=\alpha_0\sqrt{\frac{2m}{\hbar^3\omega_c}}$ and
$\beta=\beta_0\sqrt{\frac{2m}{\hbar^3\omega_c}}$.

Three exactly solvable cases are known in the literature:
\cite{Winkler}

a) When  $\beta=0$, the Rashba term couples the up-spin  level
$(\phi_{n-1},0)$ to the down-spin state $(0,\phi_n)$. The exact
eigenstates and eigenvalues of $H=H_0+H_R$ are spinors given by
\bea &&\psi^r_n=\left(\ba{c}\cos\theta_n \phi_{n-1} \\
\sin\theta_n\phi_{n}\ea \right),~~~ \psi^l_n=\left( \ba{c}
-\sin\theta_n \phi_{n-1} \\ \cos\theta_n\phi_{n} \ea \right) \nn
&&E^{r/l}_n=\hbar\omega_c
n\mp{\delta\over2}\sqrt{1+4\alpha^2n\hbar^2\omega_c^2/\delta^2}
\label{E-alpha}, \eea in which $\delta=\hbar\omega_c(1+2\xi)$.
Here $n\geq 1$ and the eigenstate $n=0$  exists only for
$\psi^l_n$ with $\theta_0=0$. The {\em mixing angle} $\theta_n$ is
given by $\tan2\theta_n=2\sqrt{n}\alpha\hbar\omega_c/\delta$ and
varies from zero (when $\alpha=0$) to $\pi/4$ (for weak magnetic
field or strong SO interaction).  Each Landau level
$\phi_{n\up\down}$ splits into two levels $\psi^l_n$ and
$\psi^r_{n+1}$ with the splitting gap $\Delta=E_n^l-E_{n+1}^r$.


b) $\alpha=0$. This time the Dresselhaus term couples
$(\phi_{n},0)$ only to the $(0,\phi_{n-1})$ leading to two new
spinor states of $H=H_0+H_D$:
 \bea
&&\psi^r_n=\left(\ba{c}\cos\theta_n \phi_{n} \\
\sin\theta_n\phi_{n-1}
\ea \right),~~~
\psi^l_n=\left(\ba{c} -\sin\theta_n \phi_{n}\\
\cos\theta_n\phi_{n-1}\ea\right) \nn
&&E^{r/l}_n=\hbar\omega_c n\mp{\delta\over2}
\sqrt{1+4\beta^2n\hbar^2\omega_c^2/\delta^2}).
\eea
Here $\delta=\hbar\omega_c(-1+2\xi)$,
$\tan2\theta_n=2\sqrt{n}\beta\hbar\omega_c/\delta$ and $n\geq 1$.
The eigenstate $n=0$  exists only for $\psi^r_n$ with $\theta_0=0$.

c) The case of  $\alpha=\pm\beta$ is in general solvable in the
absence of Zeeman coupling by going to a rotated spin coordinate
$\s^x\pm\s^y$.\cite{Loss} For $\alpha=\beta$ the SO term is simply
$H_{so}=(\alpha_0+\beta_0)(P_x+P_y)(\s^x-\s^y)$. In the diagonal
space of $\s^x-\s^y$, the eigenstates are the harmonic oscillator
levels with a spin-dependent shift of the momentum $P_{x/y}\to
P_{x/y}\pm (\alpha_0+\beta_0)m/\hbar$. For each Landau level in
the absence of Zeeman coupling, the two spin-states are
degenerate, but all states are shifted by a constant value $\Delta
E=-\hbar\omega_c(\alpha +\beta)^2/4$. Note that there is no mixing
between the two (rotated) spin states for different levels.

In the presence of Zeeman coupling there is no exact solution of
the energy spectrum. In a high magnetic field, where the Zeeman
and spin-orbit splitting are small compared to the Landau level
splitting, we can use perturbation theory near the $\alpha=\beta$
point, and up to  second order of perturbation we get $(n\ge0)$
 \bea
&&E^l_n=\hbar\omega_c\Big(n+1/2+\xi+{n\alpha^2\over1+2\xi}
-{(n+1)\beta^2\over1-2\xi}\Big)\nn
&&E^r_{n+1}=\hbar\omega_c\Big(n+1/2-\xi+{n\beta^2\over1-2\xi}
-{(n+1)\alpha^2\over1+2\xi}\Big).
\label{degen}
 \eea
When $\xi=0$ we are left with the degenerate states with a
constant shift of $-\hbar\omega_c\beta^2$.


\section{General case}
In what follows we consider the general case when both spin-orbit
terms are present. Without loss of generality we assume that
$|\beta|<|\alpha|$. Adding the Rashba interaction to the 2DEG
Landau level results in levels as in (\ref{E-alpha}). In this
case, keeping $\alpha$ constant while increasing $\beta$, it is
expected that the level splitting will decrease and the system
will return back to the levels of (\ref{degen}), when
$\beta\to\alpha$.

In the presence of both SO interaction terms, we use a Bogoliubov
transformation, and introduce new operators $c$ and $c^{\dag}$ by
\be c=(\alpha a+\ri\beta a^{\dag})/\sqrt{\alpha^2-\beta^2}
\label{Bo} \ee for the case $|\beta|<|\alpha|$. [For
$|\beta|>|\alpha|$ the definition of $c$ and $c^{\dag}$ has to be
interchanged.] This transformation is just the rescaling of the
spatial coordinates $x\to x\sqrt{{\alpha-\beta\over\alpha+\beta}}$
and $y\to y\sqrt{{\alpha+\beta\over\alpha-\beta}}$. The total
Hamiltonian transforms to
 \be
H=H_0(\omega)+H_{so}+V.\label{total}
 \ee
In this formula $H_0(\omega)$ is the Hamiltonian of a simple
harmonic oscillator with modified energy
$\omega=\omega_c(\alpha^2+\beta^2)/(\alpha^2-\beta^2)$, which
depends only on the ratio $\beta/\alpha$.  The spin-orbit term
$H_{so}$ is
\bea
H_{so}={\hbar\omega_c}\sqrt{\alpha^2-\beta^2}\left(
\begin{array}{cc}
                            0  & c \\
                           c^{\dag}  & 0
                           \end{array}\right),
 \eea
which has the form of only an effective Rashba interaction term.
Finally the spin {\em diagonal} interacting term V is
 \be
V=\hbar\omega\ri\lambda(c^2- {c^{\dag}}^2)/2,   \label{Hint}
 \ee
in which the perturbation parameter
$\lambda=2\alpha\beta/(\alpha^2+\beta^2)$ depends only on the
ratio  $\beta/\alpha$. We treat  (\ref{Hint}) as a perturbation
term. The advantage of using transformation (\ref{Bo}) lies in the
simple form of  (\ref{Hint}) which makes it possible to achieve
exact numerical solutions and most importantly a very accurate
analytical ansatz for the spectrum.


The exact eigenvalues and  eigenstates of $H_0(\omega)+H_{so}$,
which we use as the basis for the perturbation expansion, are
simply given by those in (\ref{E-alpha}), but with properly scaled
frequency $\omega$ and  SO interaction term
 \bea
&&\psi^r_n=\left(\ba{c} \cos\theta_n \phi_{n-1} \\
\sin\theta_n\phi_{n}\ea \right)
~~~\psi^l_n=\left(\ba{c}-\sin\theta_n \phi_{n-1}\\
\cos\theta_n\phi_{n} \ea \right),\nn &&E_{r/l}=\hbar\omega n
\mp{\delta\over2}\sqrt{1+4n(\alpha^2-\beta^2)\hbar^2\omega_c^2/\delta^2}\nn
&& \delta=\hbar\omega+2\hbar\omega_c\xi\nn
&&\tan2\theta_n=2\sqrt{n(\alpha^2-\beta^2)}{\hbar\omega_c/\delta}.
 \eea
Here $n\ge1$ and the eigenstate $n=0$ exists only for $\psi^l_n$
with $\theta_0=0$. Note  how the mixing angle and level splitting
are renormalized by the Dresselhaus interaction. For $\beta\to0$
we recover the results (\ref{E-alpha}). On the other hand near the
point $\alpha=\beta$ both the angle and level splitting approach
zero, as expected.

Turning to the  $V$  matrix elements, we obtain
 \be
\lambda  v_{n_in_j}^{p_ip_j}=
\langle\psi_{n_i}^{p_i}|V|\psi_{n_j}^{p_j}\rangle
\sim \delta_{n_i,n_j\pm2},~~~~~~~p_{i/j}=r,l.\label{me}
 \ee
Using the Brillouin-Wigner perturbation expansion, \cite{QM} the
basic formula of the wave-function is
 \be
|\Psi^{p_i}_{n_i}\rangle=|\psi^{p_i}_{n_i}\rangle+\sum'_{n_j,p_j}
{\lambda v_{n_in_j}^{p_ip_j}\over {\bm E}_{n_i}^{p_i}
-E_{n_j}^{p_j}}|\psi_{n_j}^{p_j}\rangle,\label{wave}
 \ee
in which the prime requires $n_i\ne n_j$ and $p_i\ne p_j$ and the
exact energy ${\bf E}_{n_i}^{p_i}$ has to be calculated from
 \be
({\bm E}_{n_i}^{p_i}-E_{n_j}^{p_j})
\langle\psi_{n_j}^{p_j}|\Psi_{n_i}^{p_i}\rangle
=\langle\psi_{n_j}^{p_j}|V|\Psi_{n_i}^{p_i}\rangle.
 \ee
The  convenience of this perturbation scheme for $V$ is that the
task of finding the wave functions can be avoided:\cite{WV} Let us start
from the second order correction to $E_{n_i}^{p_i}$:
 \be
\Delta^{(2)}(E^{p_i}_{n_i})=\lambda^2\sum'_{n_j;p_j}
{|v_{{n_i}{n_j}}^{p_ip_j}|^2
\over E^{p_i,p_j}_{n_i,n_j}}\label{delta2},
 \ee
in which $E^{p_i,p_j}_{n_i,n_j}=E_{n_i}^{p_i}-E_{n_j}^{p_j}$. For
the fourth order correction we first calculate the {\em
irreducible} term
 \be
\Delta^{(4)}_{ir}(E^{p_i}_{n_i})=\lambda^4\sum'_{n_j,n_k,n_l;p_j,p_k,p_l}
{v_{{n_i}{n_j}}^{p_ip_j}v_{{n_j}{n_k}}^{p_jp_k}
v_{{n_k}{n_l}}^{p_kp_l}v_{{n_l}{n_i}}^{p_lp_i}
\over E_{n_i,n_j}^{p_i,p_j}E_{n_i,n_k}^{p_i,p_k}E_{n_i,n_j}^{p_i,p_j}}.
 \ee
By irreducible we mean that the intermediate states are different
from the original state $\psi_{n_i}^{p_i}$. There is also another
contribution to the fourth order energy correction due to the {\em
reducible} term \be \Delta^{(4)}_{re}(E^{p_i}_{n_i})=-\lambda^4
\sum'_{n_j,n_k;p_j,p_k}
{|v_{{n_i}{n_j}}^{p_ip_j}|^2|v_{{n_i}{n_k}}^{p_ip_k}|^2
\over(E_{n_i,n_j}^{p_i,p_j})^2E_{n_i,n_k}^{p_i,p_k}}.\label{re}
 \ee
This reducible  term is the product of two irreducible terms with
complicated coefficients in general. The simple form of the
interaction term $V$ allows us to  derive (\ref{re}) just by
replacing the bare energy $E_{n_i}^{p_i}$ by the renormalized
energy $E_{n_i}^{p_i}+\Delta^{(2)}$ inside (\ref{delta2}) and
expanding  up to $\lambda^4$ terms:
 \be
\Delta^{(2)}(E^{p_i}_{n_i}+\Delta^{(2)})=\Delta^{(2)}(E^{p_i}_{n_i})
+\Delta^{(4)}_{re}(E^{p_i}_{n_i})+{\cal O}(\lambda^6).
 \ee
It can be shown  in general that all reducible terms of a given
order  can be derived from the irreducible terms of the lower
orders if we use the renormalized energy in them. That means that
we need to calculate only the irreducible terms. Once calculated,
we renormalize the energy and iterate until we get the desired
accuracy. Moreover, the calculation of the irreducible terms is
simplified  significantly because the transitions to intermediate
states are highly restricted, due to the peculiar form of
(\ref{me}). For a given level $n$ the energy can be calculated
with accuracy  $\lambda^{(N-n)/2}$, where $N$ is the total number
of  levels included ($\lambda<\beta/\alpha<1$).


Knowing the limits for $\beta=0$ (\ref{E-alpha}) and
$\alpha=\beta$ (\ref{degen}) of the spectrum, we can also propose
an approximate formula for the energy levels, increasingly
accurate  at high magnetic field or weak SO interactions. Both
$\omega$ and $\lambda$ depend on $\beta/\alpha$ and only $H_{so}$
in (\ref{total}) depends on the values of $\alpha$ and $\beta$ as
well. Rescaling back (\ref{Bo}), we notice that the effect of $V$
is to  transform  $H_0(\omega)+V$ into the simple harmonic
oscillator, with modified energy
$\omega/\sqrt{1-\lambda^2}=\omega_c$, over the whole range of
$0\leq \lambda\le1$. The main effect of $V$ is to rescale the
energy in (\ref{total}), but  this consideration  misses the
constant shift of the energy levels near the point $\beta=\alpha$.
To compensate, we add an ad hoc energy shift which gives the
correct level spectrum at $\alpha=\beta$ and at the same time does
not affect the correct limit of the formula  (\ref{E-alpha}). We
thus obtain
 \bea
&&E^{l}_n=\hbar\omega_c n
+{1\over2}\sqrt{\delta^2+4n(\alpha^2-\beta^2)\hbar^2\omega_c^2}\nn
&&+\hbar\omega_c\beta^2\big({n\over1+2\xi}-{n+1\over1-2\xi}\big)\nn
&&E^{r}_n=\hbar\omega_c n
-{1\over2}\sqrt{\delta^2+4n(\alpha^2-\beta^2)\hbar^2\omega_c^2}\nn
&&+\hbar\omega_c\beta^2\big({n-1\over1-2\xi}-{n\over1+2\xi}\big)
\label{ansats}
 \eea
in which $\delta=\hbar\omega_c(1+2\xi)$, $n\ge0$ for $E^l_n$ and
$n\ge1$ for $E^r_n$. As we will see in the examples, the agreement
of (\ref{ansats}) with the exact numerical results is excellent,
especially at strong magnetic fields, when $\alpha$ and $\beta$
are smaller.

\begin{figure}
\resizebox{.5\textwidth}{!}{\rotatebox{0}{\includegraphics{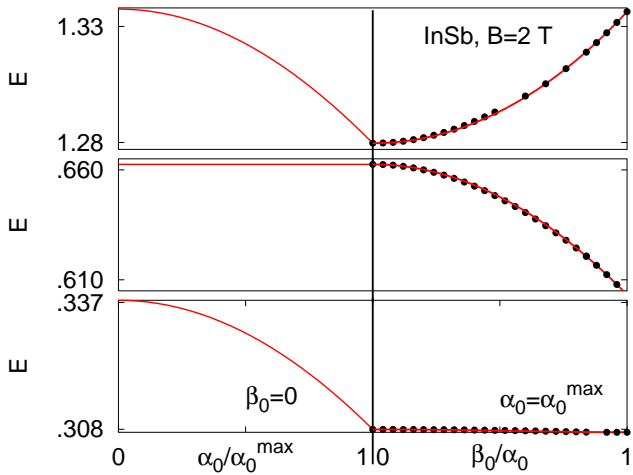}}}
\caption{(Color online) Ground state and  first two excited states
for InSb as a function of $\alpha_0/\alpha_0^{max}$ at $\beta_0=0$
(left panels) and as a function of  $\beta_0/\alpha_0$ at
$\alpha_0=\alpha_0^{max}$ (right panels). The fixed parameters are
$\alpha_0^{max}=250~meV$ \AA, $ g/g_0=-51$, $m/m_e=0.014$, $B=2
T$. Energy is in units of $\hbar\omega_c=16.5 meV$.}\label{InSb}
\end{figure}

\section{Results and Discussion}

In Fig.\ \ref{InSb} the ground state and the first two excited
states of a 2DEG in $InSb$ are plotted. The magnetic field is
chosen ($B=2 T$) such that the effect of spin-orbit splitting is
visible and at the same time let the reader appreciate the
accuracy of the approximate result (\ref{ansats}). In the left
part of this figure, first the parameter $\alpha_0$ varies from
zero to  maximum value of $\alpha_0^{max}=250 meV$ \mbox{\AA}
while the Dresselhaus parameter is zero. Here the exact spectrum
(\ref{E-alpha}) is plotted versus the Rashba interaction, akin to
an applied gate voltage. Then keeping $\alpha_0$  constant, on the
right panels of the figure, we increase the value of $\beta_0$
from zero to $\beta_0=\alpha_0$. Here the circles show the exact
numerical result, while the solid lines correspond to the
approximate solution (\ref{ansats}). Note that as $\beta$ can not
affect the ground state energy directly, the  correction to this
level starts as $\alpha^2\beta^2$, while for all other levels, the
correction is of  order   $\beta^2$ and higher.

\begin{figure}
\resizebox{.5\textwidth}{!}{\rotatebox{0}
{\includegraphics{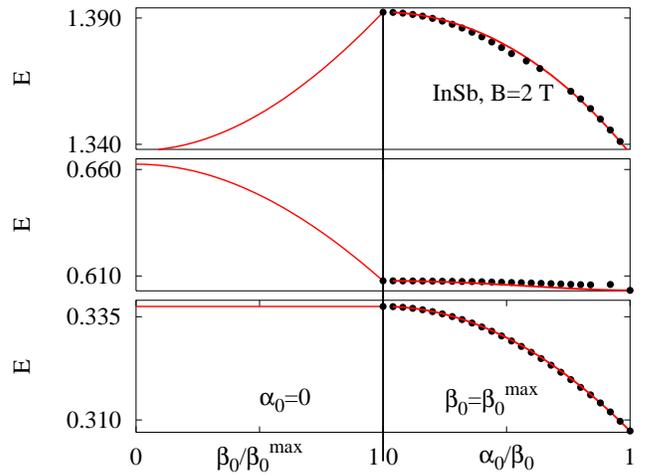}}} \caption{(Color online) Ground
state and  first two excited states in InSb as a function of
$\beta_0/\beta_0^{max}$ at $\alpha_0=0$ (left panels) and as a
function of $\alpha_0/\beta_0$ at $\beta_0=\beta_0^{max}$ (right
panels). The fixed parameters are $\beta_0^{max}=250 meV$ \AA,
$g/g_0=-51$, $m/m_e=0.014$, $B=2 T$. Energy is in units of
$\hbar\omega_c=16.5 meV$}\label{InSbr}
\end{figure}

In Fig.\ \ref{InSbr} we first switch on the Dresselhaus term (left
panels) and then keeping $\beta$ constant, we increase the value
of the Rashba interaction (right panels). This time it is the
correction to the first excited level that is of fourth order
$\alpha^2\beta^2$.

\begin{figure}
\resizebox{.5\textwidth}{!}{\rotatebox{0}{\includegraphics{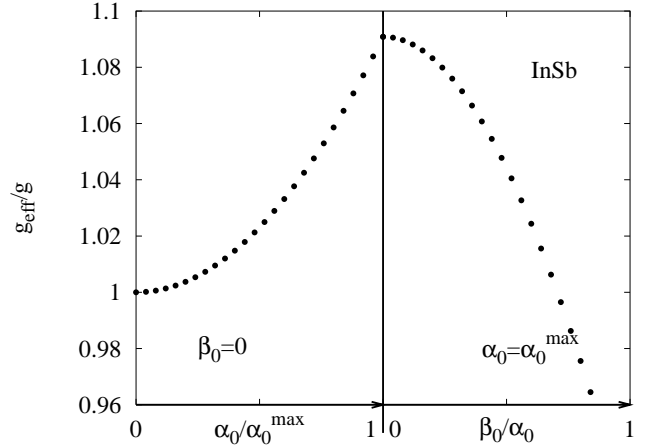}}}
\caption{(Color online) The  effective Zeeman coupling of the lowest spin-split
levels in InSb, corresponding to  Fig.\ \ref{InSb}, $B=2 T$.}
\label{InSbg}
\end{figure}

\begin{figure}
\resizebox{.5\textwidth}{!}{\rotatebox{0}
{\includegraphics{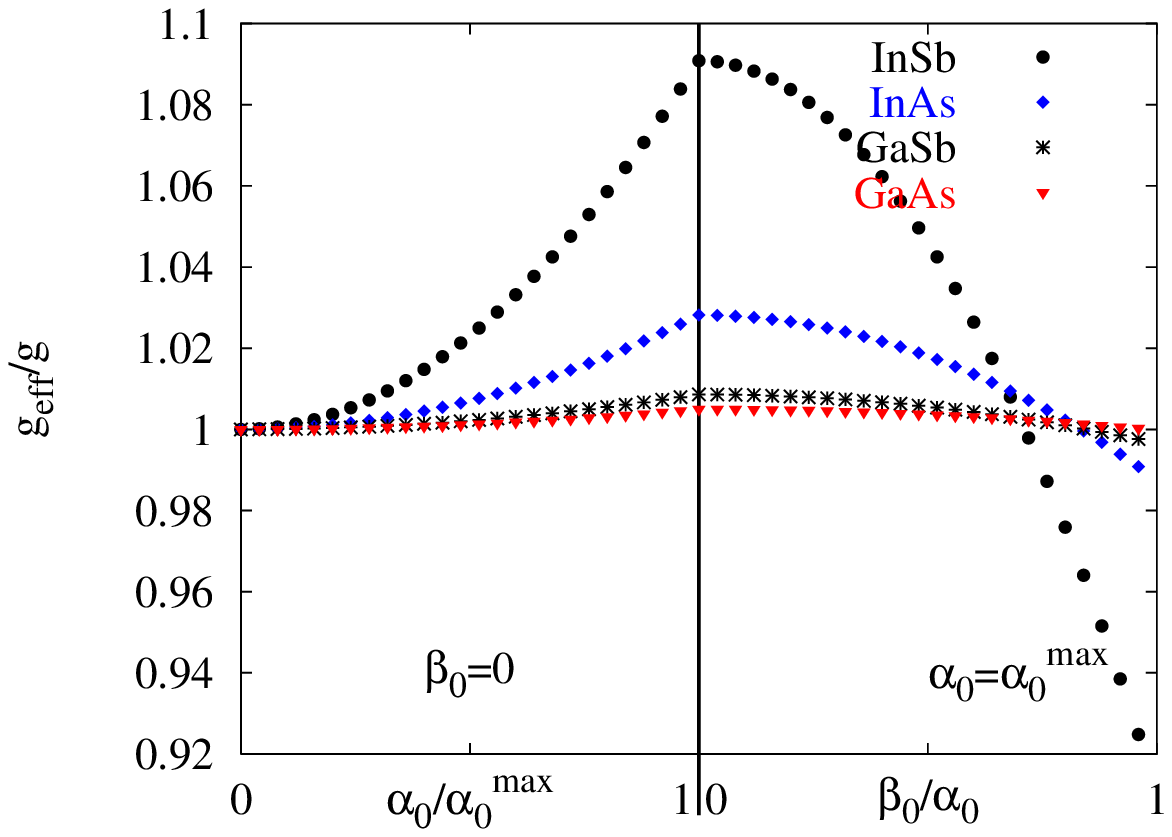}}}
\resizebox{.5\textwidth}{!}{\rotatebox{0}
{\includegraphics{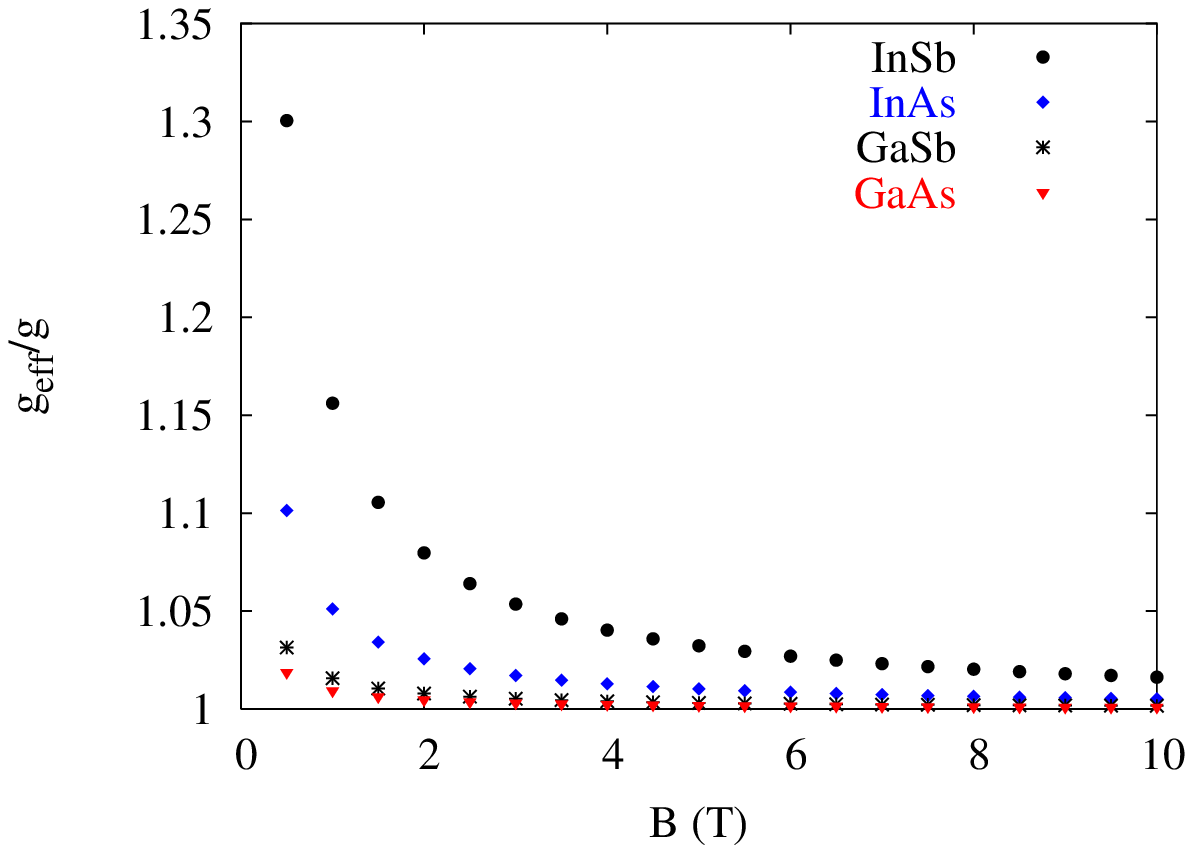}}} \caption{(Color online) Upper
panel shows the effective Zeeman coupling for different
semiconductors as a function of only $\alpha_0/\alpha_0^{max}$
(left) and as a function of $\beta_0/\alpha_0^{max}$ but in the
presence of Rashba interaction. The lower figure is the behavior
of  $g_{\rm eff}$ as a function of magnetic field for fixed value
of spin-orbit interactions.} \label{gef}
\end{figure}

Figure \ref{InSbg} shows how the SO interaction modifies the
effective Zeeman coupling. Here this parameter is defined by
$g_{\rm eff}/g=2(E^l_0-E^r_1)/(\mu_BBg)$ where $E^l_0$ and $E^r_1$
are the first excited state and the ground state respectively,
i.e, the lowest spin-split level pairs. In all cases, the Rashba
term increases $g_{\rm{eff}}$ while the Dresselhaus coupling
decreases it. The reason is simple as we note that the
$\phi_{0\up}$ state is coupled to $\phi_{1\down}$ through Rashba
interaction and pushed further down, resulting in increasing spin
splitting of each Landau level.  Contrary to this the Dresselhaus
interaction reduces the energy gap between the two spin states of
each Landau level. Although not realized for the parameters chosen
here, when the Zeeman splitting is small enough (as in low
magnetic field), the Dresselhaus term can dominate the Zeeman
coupling and interchange the ground state $\phi_{0\up}$ with
$\phi_{0\down}$. This is also found in the spectrum of quantum
dots. \cite{Carlus2}

In general the effective Zeeman coupling depends on the strength
of the spin-orbit coupling, bare Zeeman coupling and  magnetic
field. In the upper panel of Fig.\ \ref{gef} we have plotted this quantity for
different semiconductors. The value of the magnetic fields  is
fixed at $B=2 T$ for all cases. The bare value of the Zeeman
splitting and the maximum value of the spin-orbit term  is
different and chosen as in the literature .\cite{Sarma,Carlus}
The larger the spin-orbit interaction, the larger is the change in
the $g_{{\rm eff}}$.  Notice that
this modulation of $g_{{\rm eff}}$ via the spin orbit effect is in
principle controllable, via external gate  voltage (to control
$\alpha_0$). This behavior may be useful in experiments with
spins.\cite{Potok} In the lower panel of Fig.\ \ref{gef} we
plot $g_{{\rm eff}}$  as a function of the magnetic field when the
 value of the Rashba and Dresselhaus
interactions is fixed at $\alpha_0^{max}$ and
$\beta_0=\alpha_0^{max}/4$, respectively. By increasing the
magnetic field the effective Zeeman coupling decreases because
the ratio
of the spin-orbit splitting to the Landau level separation
decreases.

Once the energy is known the wave function can be calculated via
(\ref{wave}) which in turn can be used to calculate transport
parameters like charge and spin conductivity.\cite{RS,Syn,WV} That
work is in progress.


We would like to thank Roberto Romo for critical reading
of the paper and useful comments. The work was supported by
CMSS at OU and NSF-NIRT.



\begin{thebibliography}{99}

\bibitem{Datta} S. Datta and B. Das, Appl. Phys. Lett {\bf 56}, 665 (1990).
\bibitem{Rashba} E. I. Rashba, Sov. Phys. Solid state {\bf 2}, 1109 (1960).
\bibitem{Miller} J. B. Miller {\em et al.}, Phys. Rev. Lett {\bf 90},
 076807 (2003).
\bibitem{Nitta} J. Nitta, T. Akazaki, H. Takayanagi, T. Enoki, Phys. Rev. Lett. {\bf 78}, 1335 (1997).
\bibitem{Dress} G. Dresselhaus, Phys. Rev. B {\bf 100}, 580 (1955).
\bibitem{Lang} M. Langenbuch, M. Suhrke and U. Rossler, Phys. Rev. B {\bf 69},
125303 (2004).
\bibitem{Heida} J. P. Heida, B. J. vanWees, J. J. Kuipers, T. M. Klapwijk, G. Borghs, Phys. Rev. B {\bf 57}, 11911
(1998); Th. Schapers {\em et al.}, J. Appl. Phys.{\bf 98}, 4324
(1998); C. M. Hu {\em et al.}, Phys. Rev. B {\bf 60}, 7736 (1999).
\bibitem{Averkiev} N. S. Averkiev and L. E. Golub, Phys. Rev. B {\bf 60},
15582 (1999).
\bibitem{Kainz} J. Kainz, U. Rossler and R. Winkler, Phys. Rev. B {\bf 68},
075322 (2003).

\bibitem{Sinova} J. Sinova {\it et al.}, Phys. Rev. Lett. {\bf 92}, 126603
(2004); J. Schliemann and D.Loss, Phys. Rev B {\bf 71},
085308 (2005).
\bibitem{Syn} N. A. Sinitsyn, E. M.  Hankiewicz, W. Teizer, J. Sinova, Phys. Rev. B {\bf 70},
081312(R) (2004).
\bibitem{Loss} J. Schliemann, J. C. Egues, D. Loss, Phys. Rev.
Lett. {\bf 90}, 146801 (2003).
\bibitem{Pikus} F. G. Pikus and G. E. Pikus, Phys. Rev. B {\bf 51}, 16928
(1995); Ch. Schierholz, R. Kursten, G. Meier, T. Matsuyama, and U.
Merkt, Phys. Status Solidi B {\bf 233}, 436 (2002).

\bibitem{Ganichev} S. D. Ganichev {\em et al.}, Phys. Rev. Lett {\bf 92},
256601 (2004).
\bibitem{Kato} Y. Kato {\em et al.}, Nature {\bf 427}, 50 (2003).
\bibitem{Hauson} R. Hanson {\em et al.}, Phys. Rev. B {\bf 70}, 241304(R)
 (2004).
\bibitem{Potok} R. M. Potok {\em et al.}, Phys. Rev. Lett {\bf 91}, 016802
(2003).

\bibitem{Winkler} R. Winkler,
{\em Spin-orbit coupling effects in two-dimensional electron and
hole systems} (Springer-Verlag, Berlin, 2003).

\bibitem{QM} G. Baym, {\em Lectures on Quantum Mechanics}
(Addison-Wesley, New York, 1990).

\bibitem{Carlus2} C. F. Destefani, S. E. Ulloa and G. E. Marques Phys. Rev.
 B {\bf 70}, 205315 (2004).

\bibitem{Sarma} R. de Sousa and S. Das Sarma, Phys. Rev. B {\bf 68},
155330 (2003).

\bibitem{Carlus} C. F. Destefani and S. E. Ulloa, Phys. Rev. B {\bf 71},
161303(R) (2005).

\bibitem{RS}E. I. Rashba, Phys. Rev. B {\bf 70}, 201309(R) (2004).

\bibitem{WV} X. F. Wang and P. Vasilopoulos,
cond-mat/0501214.

\end{thebibliography}
\end{document}